\documentclass[letterpaper, 10 pt, conference]{ieeeconf}

\IEEEoverridecommandlockouts                              % This command is only needed if 
\usepackage{graphicx} % for pdf, bitmapped graphics files
\usepackage{mathptmx} % assumes new font selection scheme installed
\usepackage{bm} % for bold math symbols
\usepackage{times} % assumes new font selection scheme installed
\usepackage{amsmath} % assumes amsmath package installed
\usepackage{amssymb}  % assumes amsmath package installed
\usepackage{color}
\usepackage{comment}
\usepackage{booktabs}
\usepackage{colortbl}
\usepackage{cellspace}
\usepackage{multirow}

\usepackage{flushend}

\newcommand{\grayRow}{\rowcolor[rgb]{0.9,0.9,0.9}}

\let\labelindent\relax
\usepackage{enumitem}

\numberwithin{example}{section}

\title{\LARGE \bf Tube-based robust nonlinear model predictive control \\ of anaerobic co-digestion}

\author{Davide Carecci$^{1}$, Laurent Dewasme$^{2}$, Alessio La Bella$^{1}$, Gianni Ferretti$^{1}$ and Alain Vande Wouwer$^{2}$% <-this  stops a space
\thanks{$^{1}$D. Carecci (corresponding author), A. La Bella and G. Ferretti are with the DEIB, Politecnico di Milano, Italy,
        {\tt\small \{davide.carecci,alessio.labella,gianni.ferretti\} @polimi.it}}%
\thanks{$^{2}$L. Dewasme and A. Vande Wouwer are with the SECO Group, University of Mons, Belgium,
        {\tt\small \{laurent.dewasme,alain.vandewouwer\}@umons.ac.be}}%
}

\begin{document}

\maketitle
\thispagestyle{empty}
\pagestyle{empty}

\begin{abstract}
To match the growing demand for bio-methane production, anaerobic digesters need to embrace the co-digestion of different feedstocks; in addition, to improve the techno-economic performance, an optimal and time-varying adaptation of the input diet is required. These operation modes constitute a very hard challenge for the limited instrumentation and control equipment typically installed aboard full-scale plants. A model-based predictive approach may be able to handle such control problem, but the identification of reliable predictive models is limited by the low information content typical of the data available from full-scale plants' operations, which entail high parametric uncertainty. In this work, the application of a tube-based robust nonlinear model predictive control (NMPC) is proposed to regulate bio-methane production over a period of diet change in time, while warranting safe operation and dealing with uncertainties. In view of its upcoming validation on a true small pilot-scale plant, the NMPC capabilities are assessed via numerical simulations designed to resemble as much as possible the experimental setup, along with some practical final considerations.
\end{abstract}

\section{Introduction}
\label{sec:Introduction}
% Presentation of the process
To achieve energy independence and environmental sustainability, European bio-methane production is expected to increase over the incoming decades. Anaerobic digestion (AD) is a mature process in which the biodegradable components of organic wastes and biomass by-products are converted into bio-methane and biogenic carbon dioxide (biogas), and bio-fertilizer (digestate), thus helping to reduce the carbon footprint of the agricultural and waste sectors. The techno-economic competitivity of full-scale AD plants depends nowadays on the capability of the reactors to embrace a time-varying co-digestion (more than one type of feedstock) diet, and this holds true particularly for agro-zootechnical digesters. AD entails a complex cascade of slow (characteristic time of response in the order of hours-days) and non-linear biochemical processes (hydrolysis (rate-limiting step), acidogenesis, acetogenesis, methanogenesis) and requires strict control of the operating conditions: the goals are usually the maximization of bio-methane production, while keeping safe/stable operations and meeting some environmental regulations on the effluent digestate composition (e.g. total ammoniacal nitrogen (TAN) and chemical oxygen demand (COD)). Safe operations, especially during diet changes in time, are threatened by the accumulation of some intermediate products, first and foremost volatile fatty acids (VFA): the consequence of high VFA concentrations can result in the inhibition of methanogens, leading to reactor instability and/or process failure. This usually leads plant managers to opt for conservative loading conditions, thus far from the maximum production potential.

% Lack of literature in control
Different control strategies have been proposed in literature \cite{jimenez2015instrumentation}, but usually mono-digestion and/or fixed operation/set-points are considered, as well as unrealistic assumptions on the \emph{online} (i.e. automatically recorded at high frequency) data (e.g. VFA, biodegradable COD) and/or control action availability at full-scale (e.g. alkali solution) \cite{AHMED2020115599, zhou2020feeding}.
% Conditions of control at full-scale plants
Nowadays the most practical and economical control variable is the flow rate of each co-feedstock and/or the overall load (i.e. dilution rate), and it is reasonable to assume that only the measurements of biogas flow rate and composition are available \emph{online}, with the addition of some spot/low-frequency and manual TAN, VFA and total COD measurements.

% The issue of model-based control
It follows that model predictive control (MPC) is far from trivial, yet desirable for its ability to minimize a cost function while enforcing state/output/input constraints. However, a common issue of the state-of-the-art first-principle models (e.g., ADM1 \cite{batstone2002iwa}) is the presence of non-measurable states and highly uncertain parameters, due to low practical identifiability dictated by poorly informative full-scale data \cite{catenacci2024}. MPC predictions shall thus strictly rely on the performance of a state observer \cite{Dewasme2019}. To complicate matters even furher, embedding process non-linearities in the model (NMPC) is highly recommended for fairly accurate predictions, especially during diet changes.

% Carecci et al. 2024 deals with these conditions without relying on models
For these reasons, the approach of \cite{carecci2024selector} relied on 'conventional' PI control (with override strategy) and the exploitation of the correlation between the carbon dioxide over methane biogas composition ratio ($CO_2/CH_4$) and total VFA: in such work, modeling difficulties were limited to the \emph{offline} identification of an ADM1-like high-fidelity model for the definition of techno-economically \emph{offline} optimal references (i.e. real-time optimization) to be tracked with the lower-level PIs, in a 'top-down' scheme. The results of \cite{carecci2024selector}, reported in \cite{carecci2025modelling}, were partially satisfactory, but with room for improvement. 

% Other literature on model-based control
In \cite{GHANAVATI202187} the same override structure was considered, but substituting the PIs with linear adaptive MPCs, and adjusting the set-points with a fuzzy governor: however, \emph{online}-available VFA data were considered and neither the problem of diet change in time nor kinetic parameter uncertainty (the most common in literature \cite{weinrich&nelles}) were assessed. Other interesting examples of model-based control in literature can be found in \cite{mauky2016model, AHMED2020115599, KIL201763, diaz, cortes}, but always with at least one issue preventing their direct application to the current case study (mainly the limited focus to mono-digestion and/or unpractical \emph{online} measurement availability, as mentioned above).

% The tube-based NMPC: why do I want robustness? 
The coupling of uncertainties and non-linearities may degrade the performances of \emph{classical} MPC strategies, leading to the violation of the constraints (or infeasibility of the solution under constraints) and, in turn, sub-optimal or unsafe operations. A robust formulation is therefore advised, among which tube-based MPC stands out as a computationally efficient approach for handling bounded uncertainties while maintaining recursive feasibility and constraint satisfaction \cite{mayne2011, Dewasme2024}.
In this work, the 'tube' approach was applied for the first time in literature to the anaerobic digestion process: a robust NMPC was designed based on the reduced-order first-principle model developed in \cite{carecci2025modelling}, to tackle the same control problem of \cite{carecci2024selector} over the same realistic operative conditions, and to assess by means of exhaustive simulation analyses whether: (i) a model-based strategy can provide improved results compared to \cite{carecci2024selector}, and (ii) a tube-based NMPC formulation can provide significantly improved robustness with respect to a \emph{classical} one.

% Describe sections
The process models are described in Section \ref{sec:ProcessModel}, whereas Section \ref{sec:NMPCdesign} formulates the NMPC policies. Section \ref{sec:application} practically elucidates the context of the current case study and describes the numerical tests that were conducted. The results are reported in Section \ref{sec:Results}. Eventually, some conclusions and prospects are reported in Section \ref{sec:Conclusions}.

\begin{figure*}[htb]
 \begin{center}
  \includegraphics[width=0.8\textwidth]{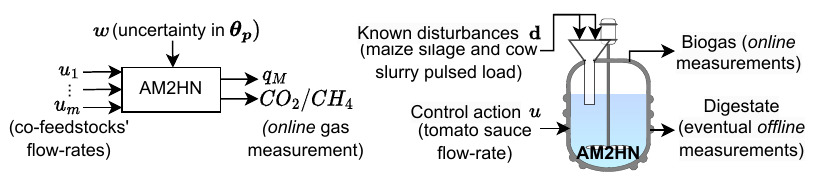}
 \end{center}
 \vspace{-0.3cm}
 \caption{Input-output schemes of the AM2HN model (on the left) and of the anaerobic co-digestion process (on the right). The latter is contextualized to the small pilot-scale operating conditions of \cite{carecci2024selector} i.e. the simulating conditions of the current work (see Section \ref{sec:application}).}
 \label{fig:scheme_process}
 \vspace{-0.2cm}
\end{figure*}

\section{Process model}
\label{sec:ProcessModel}
% {\color{red} In the following, $\mathbf{bold}$ formatting refers to vectors in the variables' space, whereas the time dependency is implicit: when relevant, time indexing is explicitly reported as subscript.} 
Overall, two first-principle and grey-box (uncertain parameters) process models were used in the control scheme: 

(a) \textbf{High-fidelity model}. The agri-AcoDM~\cite{carecci2024} was considered as an extension of the ADM1, from the mono-digestion of sewage sludge to the co-digestion of agro-zootechnical feedstocks. It consists of a strongly non-linear differential algebraic equation (DAE) system with 43 state variables.
The uncertainty in the model may lay in general both in the input parameters (i.e. the characterization of the $i^{th}$ input co-feedstock in terms of state variables' concentration, for each of the \emph{m} co-feedstocks ($\bm{\theta}_{\bm{u_i}}$)) and in stoichiometric, physico-chemical and kinetic parameters ($\bm{\theta_p}$). Kinetics is considered to be the most relevant source of uncertainty \cite{weinrich&nelles}, as a good characterization for each co-feedstock is usually available in full-scale plant (yet with low frequency, e.g. seasonal, to follow, coherently, the most relevant variations).

(b) \textbf{Reduced-order model}. An extension of the AM2HN~\cite{hassam2015} -- which in turn is an extension of the AM2~\cite{bernard2001advanced} -- to accommodate for the hydrolysis of the different \emph{m} co-feedstocks was considered~\cite{carecci2025modelling}. It traces the main dynamics of the agri-AcoDM to describe the relation between almost the same input and output vectors, but with reduced complexity by lumping some state variables (e.g. 7 bacterial populations lumped to 2). Applying mass balances to the chemical reactions included in the model, a set of ordinary differential equations (ODE, Eq.(\ref{eq:ode})) is obtained to describe the dynamics of the biochemical species concentrations (g L$^{-1}$, or mol L$^{-1}$); in compact form:
\begin{equation}\label{eq:ode}
\left\{
\begin{aligned}
   \dot{\mathbf{x}} &= \mathbf{f}(\mathbf{x},\mathbf{u},\bm{\theta}) = D \left(\mathbf{x_{in}}-\mathbf{x}\right)+ {\mathbf{K_G}} \mathbf{r}\left(\mathbf{x},\bm{\theta_p} \right) \\
    \mathbf{x_{in}} &= \frac{\sum_{i=1}^{m}u_i \bm{\theta}_{\bm{u_i}}}{\sum_{i=1}^{m}u_i}; \ D = \frac{\sum_{i=1}^{m}u_{i}}{V} \\
    \mathbf{y} &= \mathbf{g}(\mathbf{x},\bm{\theta_p})
\end{aligned}
\right.
\end{equation}
with non-linear $\mathbf{f}$ and $\mathbf{g}$, where the inputs are the flow rates (L d$^{-1}$) of the $m$ co-feedstocks ($\mathbf{u} \in \mathbb{R}^{m}$), and where the states are (i) $m$ concentrations of the biodegradable volatile solids' fraction of the co-feedstocks (hydrolysis), (ii) two bacterial populations and their -- intermediate -- growth substrates (acidogenesis and methanogenesis), (iii) dissolved inorganic carbon and (iv) total alkalinity (i.e. $\mathbf{x} \in \mathbb{R}^{n}$ with $n=m+6$). The parameters' vector $\bm{\theta}$ contains both $\bm{\theta}_{\bm{u_i}}$ and $\bm{\theta_p}$, $V$ (L) is the reactor volume, $D$ (d$^{-1}$) is the dilution rate, ${\mathbf{K_G}}$ is the stoichiometric matrix and $\mathbf{r}$ (g L$^{-1}$ d$^{-1}$, or mol L$^{-1}$ d$^{-1}$) the process kinetic rates.
A simple input-output scheme of the process and model is sketched in Figure \ref{fig:scheme_process}.
The presence of multiplicative Monod, Haldane and/or non-competitive inhibition functions in the expression of the kinetic growth rates of bacteria builds up the non-linearity of the model: for example, an Haldane-like function ($\mu_2$ (d$^{-1}$), Eq.(\ref{eq:mu2})) is used in the description of the bio-methane's production rate (output $q_M$ (mmol L$^{-1}$d$^{-1}$), Eq.(\ref{eq:qM})) from the uptake of VFA (state $S_2$ (mmol L$^{-1}$)) by methanogens (state $X_2$ (g L$^{-1}$)).
\begin{align}
\mu_2 &= \mu_{max,2}\frac{S_2}{S_2 + K_{s,2} + \frac{S_2^2}{K_{I,2}}} \label{eq:mu2} \\
q_M &= k_6 \mu_2 X_2 \label{eq:qM}
\end{align}
In Eq.s (\ref{eq:mu2})-(\ref{eq:qM}), $\mu_{\max,2}$ (d$^{-1}$) is the maximum methanogens' growth constant, whereas $K_{s,2}$ and $K_{I,2}$ (g L$^{-1}$) are the half-saturation and the inhibition constants for the nutrient $S_2$ (i.e. VFA) respectively (all three are kinetic parameters $\in$ $\bm{\theta_p}$). Such parameters are uncertain and not always practically identifiable from the available data (high collinearity and low sensitivity, especially for $K_{I,2}$), but their estimation is crucial as they define the shape of a non-monotonic static function that has a maximum $\mu_2$ value between a \emph{limited} and an \emph{inhibited} operating region.
If the co-feedstocks are nitrogen (N)-rich (e.g. animal slurries), free ammonia inhibition can be relevant and it is commonly modeled as a multiplicative non-competitive inhibition function in the $\mu_2$ expression: however, it was neglected in this work because (i) the system is highly buffered, and thus the N and pH dynamics are dampened, and (ii) the feeding rates of N-rich co-feedstocks are not used as control actions (see \cite{carecci2024} and Section \ref{sec:application}).
% Outputs
In view of full-scale applicability, the \emph{online} measurable outputs are $\mathbf{y} \in \mathbb{R}^{p}$ = [$q_M$, $CO_2/CH_4$] only (i.e. $p=2$).

The AM2HN was used to enforce the state dynamic constraints in the NMPC. Although the acido/methanogenesis core of the original AM2 model was proved to be structurally locally observable even with gas data only \cite{Dewasme2019}, an increase of the model dimensionality was needed to obtain fairly good prediction ability in time-varying co-digestion operations. Since the majority of the model states are not measurable, the design of a state observer would be required in practice, preempted by the assessment of the observability property of the model, but this aspect goes beyond the scope of the present manuscript and will be tackled in oncoming works.

% What high-fidelity used for
After the parameter estimation carried out in~\cite{carecci2025modelling}, the agri-AcoDM was used: (i) to perform an \emph{offline} constrained optimization of the diet (see \cite{carecci2024} and Section \ref{sec:application}) that defined the reference output trajectories ($\mathbf{y_{ref}}$), and (ii) to generate in-silico informative data for the identification of the AM2HN's $\bm{\theta_p}$ (then refined on real data after parameter subset selection (PSS))~\cite{carecci2025modelling}. 
\section{NMPC design}
\label{sec:NMPCdesign}
% Control objectives
The control objectives are thus to (i) track $\mathbf{y_{ref}} \in \mathbb{R}^p$ while (ii) warranting safe operations i.e. keeping some outputs/states below expert-defined thresholds. As a result, the goal was to solve a multivariable and multi-objective control problem with nonlinear dynamics, partial observability, parametric uncertainty, constraints and time-varying setpoints. 
In its most simple formulation, the objective function $J$ to be minimized reads:
\begin{equation} \label{eq:phi}
J = \sum_{t=t_0}^{t_0+H_p-1} \Phi_t, \ \text{where} \ \Phi_t = \left\| \frac{\mathbf{y}_{t} - \mathbf{y_{ref}}_{t}}{\mathbf{\Bar{y}}} \right\|_{2,\mathbf{W_y}}
\end{equation}
where $\mathbf{W_y}$ is a diagonal square matrix ($\in \mathbb{R}^{p \times p}$) that sets the output-specific weights of the 2-norm, $t_0$ is the time index of the current control step and $H_p$ is the finite prediction horizon of the NMPC. The constant vector $\mathbf{\Bar{y}}$ normalizes element-wise the different outputs' tracking errors (e.g. with the time averages of the data already available at $t=0$ for each output). 

% Presentation of the tube
With respect to a \emph{classical} NMPC, the standard tube-based formulation roots in the stabilizing MPC approach, for which the control problem is split into a \emph{nominal} open-loop and an \emph{ancillary} 'lower-level'/closed-loop problem: the latter, subject to disturbances (of the real system), tries to follow the solutions provided by the \emph{nominal} undisturbed problem applying an additional error feedback \cite{mayne2011}. Recalling the notation of \cite{Dewasme2024}, the \emph{nominal} (states $\mathbf{z}$ and inputs $\bm{\nu}$) and the real disturbance-affected (states $\mathbf{x}$ and inputs $\mathbf{u}$) systems are defined by Eq.(\ref{eq:zdot}) and Eq.(\ref{eq:xdot}) respectively. 
\begin{align}
    \mathbf{\dot{z}} &= \mathbf{f}(\mathbf{z}, \bm{\nu}) \tag{5a} \label{eq:zdot} \\
    \mathbf{\dot{x}} &= \mathbf{f}(\mathbf{x}, \mathbf{u}) + \mathbf{w} \tag{5b} \label{eq:xdot}
\end{align}
where $\mathbf{f}$ is the same mapping of Eq.(\ref{eq:ode}) and $\mathbf{w}$ are bounded unknown disturbances such as $\mathbf{w} \in \mathbb{W}$, a convex set that contains the origin. This standard tube-based approach performs well if an invariant set of the disturbed system is known 'a priori'. Since this cannot be completely assumed to hold in our case, due to model uncertainties, an additional degree of freedom was introduced: by updating recursively the \emph{nominal} solution, disturbances are taken into account over a certain extent and a certain degree of feedback is introduced in the \emph{nominal} problem \cite{BAYERALLGOWER}, reducing the effort of the \emph{ancillary} controller. Hereafter, the standard and this latter approaches are referred to as '\emph{offline}-' and '\emph{online}-tube' respectively. The \emph{online}-tube formulation of the NMPC was derived following the guidelines of \cite{Dewasme2024}: its \emph{nominal} and \emph{ancillary} problems are thus descibed by the systems of Eq.s (\ref{eq:6a})-(\ref{eq:6g}) and (\ref{eq:7a})-(\ref{eq:7h}) respectively:
\begin{align} 
    \min_{\bm{\nu}} \quad & J  \tag{6a} \label{eq:6a} \\
    \text{s.t.} \quad & \mathbf{\dot{z}} = \mathbf{f}(\mathbf{z}, \bm{\nu}) \tag{6b} \label{eq:6b} \\
    & \mathbf{y} = \mathbf{g}(\mathbf{z}, \bm{\nu}) \tag{6c} \label{eq:6c} \\
    & \bm{\nu}_t = \bm{\nu}_{t_{0+H_c-1}}, \quad t \in [t_{0+H_c}, t_{0+H_p-1}] \tag{6d} \label{eq:6d} \\
    & \bm{\nu_{\textbf{lb}}} \leq \bm{\nu}_t \leq \bm{\nu_{\textbf{ub}}}, \quad t \in [t_0, t_{0+H_p-1}] \tag{6e} \label{eq:6e} \\
    & \Delta \bm{\nu_{\textbf{lb}}} \leq \Delta \bm{\nu}_t \leq \Delta \bm{\nu_{\textbf{ub}}}, \quad t \in [t_0, t_{0+H_c-2}] \tag{6f} \label{eq:6f} \\
    & \mathbf{z_{lb}} \leq \mathbf{z}_t \leq \mathbf{z_{ub}}, \quad t \in [t_0, t_{0+H_p-1}] \tag{6g} \label{eq:6g} \\
    \notag \\
    \min_{\mathbf{u}, \ \mathbf{z}_{t_0}^*} \quad & \bigl[ w_{y_{H_p}} \Phi_{t_{0+H_p}} + \notag \\
    & + \sum_{t=t_0}^{t_0+H_p-1} \left( w_x \left\| \mathbf{x}_t - \mathbf{z}_t^* \right\|_2 + w_u \left\| \mathbf{u}_t - \bm{\nu}^*_t \right\|_2 \right) \bigl] \tag{7a} \label{eq:7a} \\
    \text{s.t.} \quad & \mathbf{\dot{z}^*} = \mathbf{f}(\mathbf{z}^*, \bm{\nu}^*), \quad \mathbf{z}^* \in \mathbb{Z}, \quad \bm{\nu}^* \in \mathbb{V}  \tag{7b} \label{eq:7b} \\
    & \mathbf{\dot{x}} = \mathbf{f}(\mathbf{x}, \mathbf{u}) \tag{7c} \label{eq:7c} \\
    & \mathbf{y} = \mathbf{g}(\mathbf{x}, \mathbf{u}) \tag{7d} \label{eq:7d} \\
    & \mathbf{u}_t = \mathbf{u}_{t_{0+H_c-1}}, \quad t \in [t_{0+H_c}, t_{0+H_p-1}] \tag{7e} \label{eq:7e} \\
    & \mathbf{u_{lb}} \leq \mathbf{u}_t \leq \mathbf{u_{ub}}, \quad t \in [t_0, t_{0+H_p-1}] \tag{7f} \label{eq:7f} \\
    & \Delta \mathbf{u_{lb}} \leq \Delta \mathbf{u}_t \leq \Delta \mathbf{u_{ub}}, \quad t \in [t_0, t_{0+H_c-2}] \tag{7g} \label{eq:7g} \\
    & \mathbf{x_{lb}} \leq \mathbf{x}_t \leq \mathbf{x_{ub}}, \quad t \in [t_0, t_{0+H_p-1}] \tag{7h} \label{eq:7h}
\end{align}
where $H_c$ and $H_p$ are the control and the prediction horizons respectively (with $H_c < H_p$ to reduce the computational time), and the $\mathbf{lb}$ and $\mathbf{ub}$ subscripts refer to the lower- and upper-bounds that define the feasible sets $\mathbb{X}$, $\mathbb{U}$ and $\mathbb{DU}$ of $\mathbf{x} \in \mathbb{R}^{n}, \mathbf{u} \in \mathbb{R}^{m \times H_p}$ and $\Delta_{\mathbf{u}} \in \mathbb{R}^{m \times (H_p-1)}$ respectively. $w_x$ and $w_u$ are the weights applied to the state and input distances expressed as Euclidean norms, respectively, and $w_{y_{H_p}}$ is the weight of the terminal cost $\Phi_{t_{0+H_p}}$. $\mathbb{Z} \subseteq \mathbb{X}$ and $\mathbb{V} \subseteq \mathbb{U}$ define the feasible sets of the \emph{nominal} states and inputs ($\mathbf{z} \in \mathbb{R}^{n}$ and $\bm{\nu} \in \mathbb{R}^{m \times H_c}$) respectively. As the \emph{nominal} problem is solved first, the optimal solutions $\mathbf{z}^*$ and $\bm{\nu}^*$ are then the 'reference' trajectories tracked by the \emph{ancillary} problem. However, in the \emph{online}-tube, instead of penalizing the distance from these trajectories as they are, the presence of (\ref{eq:7b}) re-evaluates $\mathbf{z}^*$ in the \emph{ancillary} problem, applying $\bm{\nu}^*$ to the model and starting from the additional decision variable $\mathbf{z}_{t_0}^*$. The "re-optimized" $\mathbf{z}_0^*$ computed in the \emph{ancillary} problem at the $k^{th}$ control step is then used to constrain the initial states of the \emph{nominal} problem at the $k^{th}+1$ step (i.e. $\mathbf{z}_{t_0,k}$ of Eq.(\ref{eq:6b}) is set equal to $\mathbf{z}_{t_0,k-1}^*$ of Eq.(\ref{eq:7b})). The initial conditions at $k=0$ of the state trajectories computed in the \emph{nominal} problem ($\mathbf{z}_{0,0}$) can be set, for example, to the steady-state values retrieved from a previously-run open-loop simulation of the current inputs. Eventually, (\ref{eq:7c}) enforces the constraint of the model dynamics to start from the currently measured/estimated states of the real plant ($\mathbf{x}_{t_0}$ of Eq.(\ref{eq:xdot})).

% Deduce classical and offline-tube
From the system of equations reported, the reader can also infer the \emph{classical} and \emph{offline}-tube formulations, that were not reported for the sake of brevity, but that will be recalled in Sections \ref{sec:application}-\ref{sec:Results} to derive logical and consequential considerations to find the best control strategy, and to compare the robustness of the tube formulations against the \emph{classical} one. The \emph{classical} NMPC is formulated as Eq.s (\ref{eq:7a})-(\ref{eq:7h}) evaluating (\ref{eq:phi}) instead of the cost function in (\ref{eq:7a}), and removing (\ref{eq:7b}). Instead, in the \emph{offline}-tube, the cost function of the \emph{ancillary} problem remains as (\ref{eq:7a}), but $\mathbf{z}_{t_0}^*$ is not a degree of freedom anymore: in practice, (\ref{eq:7b}) disappears as the solutions of the \emph{nominal} problem (the $\mathbf{z}^*$ and $\bm{\nu}^*$ trajectories) are computed \emph{once and for all} over the whole horizon for which $\mathbf{y_{ref}}$ is given and not updated recursively (indeed they play a role 'similar' to $\mathbf{y_{ref}}$ in the \emph{ancillary} problem's cost function).

% Assumptions of Mayne
The redefinition of the main assumptions of \cite{mayne2011} in the context of the current case study is very similar to the one present in \cite{Dewasme2024}: an exhaustive Monte-Carlo analysis was conducted to verify the stability of system (\ref{eq:ode}) around the equilibrium points ($\mathbf{\Bar{x}_{ref}}, \mathbf{\Bar{u}_{ref}}$) that correspond to the steady-state values of $\mathbf{y_{ref}}$ ($\mathbf{\Bar{y}_{ref}}$). It follows that the system (\ref{eq:7a})-(\ref{eq:7g}) forces the states of the real plant to be contained in a compact tube set centered in $\mathbf{z}^*$, if the unknown disturbances are bounded. This time-varying tube set cannot always be determined in practice for general nonlinear systems, but its exact determination is not mandatory to ensure the closed-loop stability and the convergence of the state trajectories to such set: however, the size of the tube can be approximated following a Monte-Carlo analysis of the simulated closed-loop \cite{mayne2011, Dewasme2024} (see Section \ref{sec:application}).

% Slackness
Since the feasible sets of the \emph{nominal} problem are taken as subsets of the \emph{ancillary} ones, a practical advantage of the tube is also given by the possibility to 'relax' the constraints of the \emph{ancillary} problem by tightening the ones of the \emph{nominal}, mitigating the risk of recursive feasibility's issues. However, as common of many industrial applications of \emph{classical} MPCs, a similar role can be played by the introduction of 'slack' variables ($\bm{\epsilon} \in \mathbb{R}^{n}_{\geq0}$), that are intended to relax the state bound constraints by being heavily weighted in the cost function to penalize their exploitation \cite{Rawlings2000}. Among the others, some considerations on the presence of slack variables in the NMPC formulations will be reported in Section \ref{sec:Results}.

\section{Application}
\label{sec:application}
The NMPC was designed to tackle the same control problem described in \cite{carecci2024selector} and to allow for a comparison between conventional and model-based control schemes in pursuing the goals elucidated in Section \ref{sec:NMPCdesign}. 
Thus, in this study, $\mathbf{y_{ref}}$ = [$q_{M,ref}, \ (CO_2/CH_4)_{ref}$], and the quantities to be kept below expert-defined thresholds to warrant 'safe' operations are in particular $S_2$ and $CO_2/CH_4$.
The operating conditions to be resembled in the following simulations are thus the same of \cite{carecci2024selector}, in which a small pilot-scale reactor ($V= 12$ L) was operated under conditions resembling the ones of a nearby full-scale plant that co-digests maize silage, cow slurry and other industrial by-products (temperature $\sim$ 42-43°C). However, the lack of controllable screw-presses in the pilot-plant equipment imposed the need to feed manually and impulsively (3 times/week) the TS-rich co-feedstocks (maize silage and cow slurry). Tomato sauce was selected as a pumpable feedstock, substituting the organic loading rate (OLR) of the other industrial by-products: as a result, the tomato sauce flow rate was the only control action ($m$ = 1 in Section \ref{sec:NMPCdesign}), whereas the other co-feedstocks' flow rates were treated as (strong) known disturbances $\mathbf{d}$ (see Figure \ref{fig:scheme_process}). The calibrated agri-AcoDM -- implemented in OpenModelica i.e. using an open-source, high-level, declarative and object-oriented modeling language~\cite{Fritzson2020} -- was used for the \emph{offline} optimization of the diet, thus defining both the $\mathbf{y_{ref}}$ and $\mathbf{d_{ref}}$ trajectories over a simulation horizon of 44 days: $\sim$ 2 weeks in the above-mentioned 'steady-state' diet ($D \sim$ 0.031 d$^{-1}$, OLR $\sim$ 2.8 g$_{COD}$ L$^{-1}$ d$^{-1}$ of which 23\% tomato sauce, 49\% cow slurry, 27\% maize silage), $\sim$ 2 weeks of ramp transition and $\sim$ 2 weeks in the optimized 'steady-state' diet with increased maize and reduced slurry (see \cite{carecci2024selector}).%\textcolor{red}{In such single-input multiple-outputs (SIMO) context, there is not full controllability, but, since the AM2HN was trained also on agri-AcoDM in-silico data, setting the $\mathbf{d_{ref}}$ that results out the constrained agri-AcoDM \emph{offline} optimization avoids the uprising of instability at the equilibrium that corresponds to $\mathbf{y_{ref}}$}.

Different simulation tests, whose main characteristics are listed in Table \ref{tab:tests}, were conducted to assess the controllers' adequacy and robustness.

% Table generated by Excel2LaTeX from sheet 'table1'
\begin{table}[htbp]
  \centering
  \caption{Summary of testing conditions}
    \begin{tabular}{ccccc}
    \toprule
    \textbf{ID} & \textbf{Formulation} & \textbf{Slack var.} & \textbf{Real plant} & \textbf{Tight} $\mathbb{Z}$ \\
    \midrule
    \grayRow TEST0 & \textit{classical} & no & AM2HN & no \\
    TEST1 & \textit{offline-tube} & no & AM2HN & no \\
    \grayRow TEST1b & \textit{offline-tube} & no & AM2HN & no \\
    \midrule
    TEST2 & \textit{classical} & yes & AM2HN & yes \\
    \grayRow TEST3 & \textit{online-tube} & yes & AM2HN & yes \\
    \midrule
    TEST4 & \textit{classical} & yes & agri-AcoDM & yes \\
    \grayRow TEST5 & \textit{offline-tube} & yes & agri-AcoDM & yes \\
    TEST6 & \textit{online-tube} & yes & agri-AcoDM & yes \\
    \bottomrule
    \end{tabular}%
  \label{tab:tests}%
\end{table}%

% Choice of hyper-parameters
The control interval $T_c$ was set to 6 hours, based on the system's typical time of response (values within the 2-12 hours range are recommended, depending on the hydrolytic characteristics of the diet). $H_p$ was set to 10 intervals i.e. prediction up to 2.5 days ahead, considering the system's settling time and the frequency of the $\mathbf{d_{ref}}$ impulse feeding (3 times/week in the realistic tests). $H_c$ was set to 2 intervals, following the reasoning of \cite{Dewasme2024}.
$\mathbb{X}$ was set reasonably wide, and considering the process a priori knowledge ($\mathbb{X} \subseteq \mathbb{R}_{\geq0}$ and, for instance, an upper threshold on $S_2$ to avoid operations near the inhibition region of the Haldane curve), whereas $\mathbb{U}$ and $\mathbb{DU}$ were set according to the physical limitations of the actuator (peristaltic pump). 
Unless specified otherwise, $\mathbf{W_y}$ was set to the identity matrix, whereas $w_{y_{Hp}}$, $w_x$ and $w_u$ were hyper-parameters adjusted by trial-and-error for the different tests (values reported in Section \ref{sec:Results}).
The resulting constrained nonlinear program were solved in 'multiple shooting' using the ‘IPOPT’ solver embedded in the CasADi framework \cite{Andersson2019}.

\begin{figure}[htb]
 \begin{center}
  \includegraphics[width=1\columnwidth]{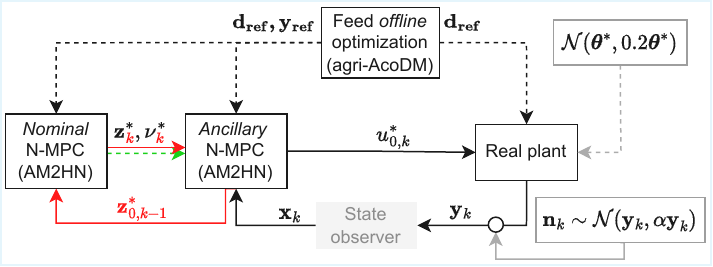}
 \end{center}
 \caption{\emph{Online}-tube NMPC block diagram contextualized in its application to TEST3 and TEST6 at the $k^{th}$ control step. The '$\sim$' symbol indicates the extraction of one random realization from the Gaussian distributions. The red/green formatting indicates the elements to be eliminated/added to derive the \emph{offline}-tube NMPC block diagram. Dashed lines indicate the \emph{offline} information exchange. The "State observer" block is actually not present in the current work, but fundamental for the practical applicability of the control scheme and it will be included in future works (see Section \ref{sec:Conclusions}).}
 \label{fig:blocks}
\end{figure}

% Uncertainties (simple tests)
To compare the different NMPC formulations, some preliminary tests (TEST0-3) were conducted with the AM2HN model as the real plant (the simulation period was reduced from 44 to 30 days to save computation time), where the unknown disturbances $\mathbf{w}$ were the uncertainties on the kinetic parameters $\bm{\theta}_{\bm{kin}} \subset \bm{\theta_p}$ = [$\mu_{max,1}, \mu_{max,2}, K_{s,1}, K_{s,2}, K_{I,2}$]: in the Monte-Carlo analysis, the closed-loop was run with 30 different realizations of $\bm{\theta}_{\bm{kin}}$, taken from Gaussian distribution centered in the \emph{expected values} ($\bm{\theta}^*$) identified in \cite{carecci2025modelling} and with 20\% relative standard deviation. Some measurements' noise $\mathbf{n}$ was also added from zero-mean Gaussian distributions with relative standard deviation of 5\% and 2\% for $q_M$ and $CO_2/CH_4$ respectively. %Call it n_t and write it explicitely in equations (8c)? 
The \emph{online}-tube (the most articulated) block diagram is shown in Figure \ref{fig:blocks}, where it is made clear (in red color) the additional degree of freedom with respect to the \emph{offline}-tube, i.e. the feedback of $\mathbf{z}_{0,k-1}^*$ from the \emph{ancillary} to the \emph{nominal} problem at each control step $k$. The diagram of the \emph{classical} NMPC can be deduced by removing the \emph{nominal} block. 

% Difference between TEST1 and TEST1b
Due to the multi-objective nature of \eqref{eq:7a}, a Pareto-type trade-off exists between tighter output, state or input tubes: considering that the tightening of the input tube is not of primary interest in the current study, TEST1b was run with a different choice of the weighting hyper-parameters with respect to TEST1, in order to report some considerations about such design choices. In particular, with respect to all the other tests (values equal to the ones of TEST1), TEST1b set much higher relevance to the tracking of the \emph{nominal} "re-optimized" state trajectories ($w_x$) than to the terminal output cost ($w_{y_{Hp}}$).
% Pareto solution
In addition, since $p=2$, \eqref{eq:phi} is bi-objective, so that the choice of \emph{diag}($\mathbf{W_y}$) is another degree of freedom that may lead to Pareto-efficient solutions or not. To preliminarily investigate over this, TEST0 and TEST1 were repeated with \emph{diag}($\mathbf{W_y}$) = \{[0,1], [0.1, 0.9], [0.25, 0.75], [0.5,0.5], [0.75, 0.25], [0.9, 0.1], [1, 0]\}.
% Slack tests
To avoid recursive feasibility issues when tightening the state constraints in the \emph{classical} and tube \emph{nominal} problems, the impact of the inclusion of slack variables was also assessed (TEST2-3).

% Realistic tests
Afterward, some more realistic tests (TEST4-6) were performed to resemble as much as possible the real small pilot-scale conditions (see Table \ref{tab:tests}), in view of the future experimental validation of the controller. Both the impulse nature of $\mathbf{d_{ref}}$ (manual feeding) and the agri-AcoDM as real plant were considered. In addition, to challenge the controller even more, a temporary unknown reduction of 60\% in the kinetic constant of the methanogens' growth ($k_{m,ac}$ (g$_{COD_s}$ g$_{COD_x}^{-1}$ d$^{-1}$)) during the \emph{transient} period from one diet to another was artificially introduced in the agri-AcoDM (see Figure \ref{fig:openloop_adm1}): this allowed the reproduction of a temporary VFA accumulation over the \emph{transient} period similar to the one occurred in the experiment described in \cite{carecci2024selector} (result data shown in \cite{carecci2025modelling}), and that was not completely seizable by an agri-AcoDM with fixed parameters. In these tests, for the Monte-Carlo analysis, the closed-loop was run with 30 different realizations of the amplitude and duration of the trapezoidal $k_{m,ac}$ reduction in time, considering Gaussian distributions centered in 60\% and 7 days, and with 7\% and 1 day as relative standard deviations respectively.

% Comparison classical vs tube conditions
To allow for a fair comparison between the \emph{classical} and the tube-based strategies, the two were challenged always with the same disturbance conditions, for each test, and the Monte-Carlo analyses were performed using the same random seeds. Moreover, the same constraints were always considered for the \emph{classical} and tube \emph{nominal} problems (i.e. $\mathbb{X}$ and $\mathbb{U}$ in \emph{classical} were equal to $\mathbb{Z}$ and $\mathbb{V}$ in \emph{nominal}). In the tube, the \emph{nominal} $\mathbb{V}$ and the \emph{ancillary} $\mathbb{U}$ were equal for all tests, whereas $\mathbb{Z}$ was tightened (by individual restrictions of the $\mathbf{lb}$ and $\mathbf{ub}$) or not in the different tests as specified in Table \ref{tab:tests}: from the experimental results reported in \cite{carecci2025modelling}, $\mathbb{Z}$ was tightened to guarantee the desired 'degree of safety' against high $S_2$ ($ub$ = 20 mmol L$^{-1}$) and low $X_2$ ($lb$ = 1 g L$^{-1}$).
% Metrices to compare the different tests
To quantify the benefits of the tube, the mean and maximum deviations from the central average path ($\Bar{\sigma}(S_2)$ and $\sigma_{max}(S_2)$, that characterize the size of the trajectory corridors) and the maximum values of $S_2$ were considered, whereas the average relative root-mean-square error over the 30 runs ($\Bar{RMSE}$) computed as in \cite{Dewasme2024} was used to compare the accuracy in $\mathbf{y_{ref}}$ tracking.

% Comparison classical NMPC vs selector PI
Eventually, considering the \emph{expected value} of the realizations of the $k_{m,ac}$ reduction in the agri-AcoDM, a comparison between a \emph{classical} NMPC run and the 'override-PIs' scheme of \cite{carecci2024} is reported. For the sake of fairness, the parameters of the 'override-PIs' were thus re-tuned (method described in \cite{carecci2024selector}) to set the same $T_c$ for both controllers. 
\section{Results}
\label{sec:Results}
\begin{figure}[t]
\begin{center}
\includegraphics[width=1\columnwidth]{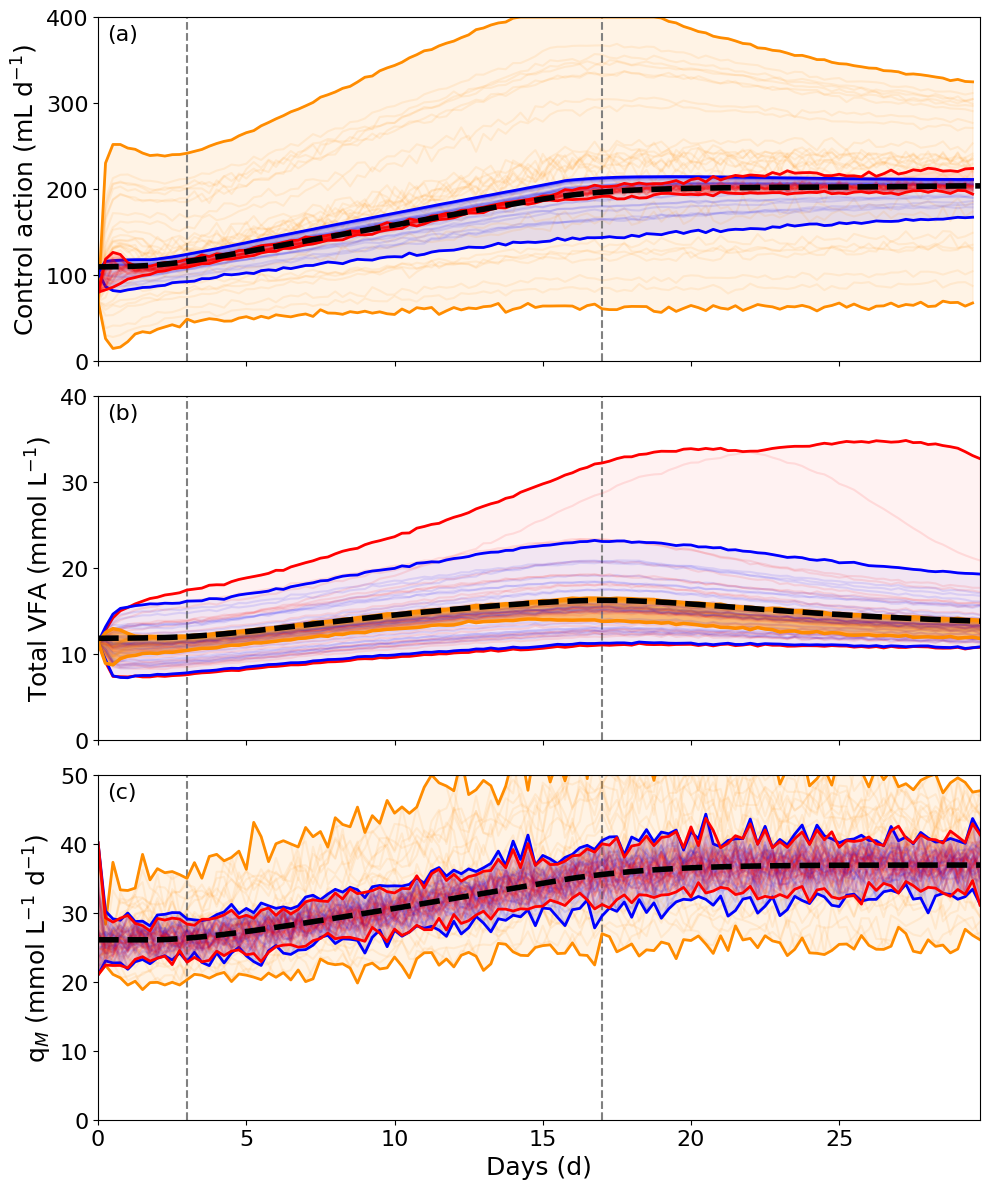}
\end{center}
\vspace{-0.3cm}
\caption{Results from the Monte-Carlo analyses of TEST0 (red), TEST1 (blue) and TEST1b (orange). The two vertical dotted gray lines delimit the time window of the $\mathbf{d_{ref}}$ ramps. The black thick dotted lines are: $v^*$ and $\mathbf{z^*}$ in (a) and (b), and $q_{M,ref}$ in (c).}
\label{fig:closedloop_simpletests}
\end{figure}

Table \ref{tab:performances} reports, for each test, the metrics to compare the different formulations of the NMPC problem.
% Simple tests
Figure \ref{fig:closedloop_simpletests} shows the trajectories of the 30 \emph{classical} (TEST0, red color) and \emph{offline}-tube (TEST1 and TEST1b, blue and orange colors) runs: to qualitatively assess the corridor tightening capability of the tube, the lower and upper bounds of the input, state and output trajectories were highlighted.

% Table generated by Excel2LaTeX from sheet 'table2'
\begin{table}[htbp]
  \centering
  \caption{Tracking performances and robustness metrics}
    \begin{tabular}{cccr}
    \toprule
    \textbf{ID} & $\mathbf{\Bar{RMSE}}$ $\mathbf{q_{M}}$; $\mathbf{CO_2/CH_4}$ & $\mathbf{\Bar{\sigma}(S_2)}$; $\mathbf{\sigma_{max}(S_2)}$ & \multicolumn{1}{c}{$\mathbf{S_{2,max}}$} \\
    \midrule
    \grayRow TEST0 & 5.4; 2.5  & 45.6; 52.7  & 34.8 \\
    TEST1 & 6.1; 2.3  & 32.4; 35.9  & 23.2 \\
    \grayRow TEST1b & 16.2; 2.1  & 7.9; 19.3  & 16.5 \\
    \midrule
    TEST2 & 5.6; 2.4  & 32.2; 39.3  & 20.4 \\
    \grayRow TEST3 & 5.6; 2.4  & 32.6; 39.6  & 20.4 \\
    \midrule
    TEST4 & 19.4; 7.4  & 9.5; 30.3  & 31.7 \\
    \grayRow TEST5 & 22.4; 8.1  & 6.0; 24.6  & 25.4 \\
    TEST6 & 19.9; 8.0  & 8.7; 27.9  & 30.1 \\
    \bottomrule
    \end{tabular}%
  \label{tab:performances}%
\end{table}%

The weights $w_{y_{H_p}}, w_x$ and $w_u$ of the \emph{ancillary} cost function were set to 90, 1, 9 and 1, 1, 0.1 for TEST1 and TEST1b respectively: comparing the results of these two designs it is made clear that a greater tightening of the state tubes can be achieved only at the expense of a poorer $q_{M,ref}$ tracking. Due to the nature of the $\bm{\theta}_{\bm{kin}}$ uncertainty, that ultimately is multiplicative in the algebraic relation between the $S_2$ state and the $q_M$ output (Eq. (\ref{eq:qM})), the tightening of both tubes is conflicting: from the results, a direct relation between the control action and the bio-methane flow rate in the model is revealed, as well as an 'inverse' relation between the total VFA concentration and the bio-methane flow rate. On the other hand, a tightening of the total VFA and the $CO_2/CH_4$ tubes is not conflicting, coherently with \cite{carecci2024selector}. 
Apparently, due to (i) the model non-linearities, (ii) the (realistic) assumptions on $\mathbf{w}$ and (iii) the model structure (a cascade of reactions), the tube-based NMPC was not able to tackle both objectives of the control problem; as typical of many engineering applications, a trade-off between robustness and tracking performance shall be found, for instance with the \emph{offline}-tube design of TEST1, in which a 35\% lower $S_{2,max}$ value is obtained losing the 12\% in $q_{M,ref}$ tracking accuracy only compared to TEST1b. 

% Pareto-solutions
When analyzing the simulations repeated with different values of $\mathbf{W_y}$, it is consistent and interesting to note that with a \emph{classical} NMPC, with respect to a robust \emph{offline}-tube formulation, the dispersion of the solutions across the bi-objective [$\Bar{RMSE}_{q_M}$, $\Bar{RMSE}_{CO_2/CH_4}$] space was significantly higher, meaning that the robust formulation is less sensitive on the tuning of $\mathbf{W_y}$.
For both TEST0 and TEST1, the equal-weights solution was found to be Pareto-efficient. For TEST0, however, considering the Euclidean distance of the outputs' $\Bar{RMSE}$ from the origin and the $S_{2,max}$ values as objectives, the equal-weight solution resulted to be slightly dominated by the solution with \emph{diag}($\mathbf{W_y}$) = [0.75, 0.25]. For TEST1, with the same objectives, the solution with [0,1] was the only one significantly dominated by the others.

% Simple tests-Introduction of slack
In any case, in light of these considerations and looking for a good trade-off design for this specific application and control objectives, a \emph{classical} NMPC with tight state constraints and slack variables may be able to cancel out the benefits of a tube-based formulation, as can be deduced from the very similar results of TEST2 and TEST3 in Table \ref{tab:performances}.

\begin{figure}[!htb]
 \begin{center}
  \includegraphics[width=1\columnwidth]{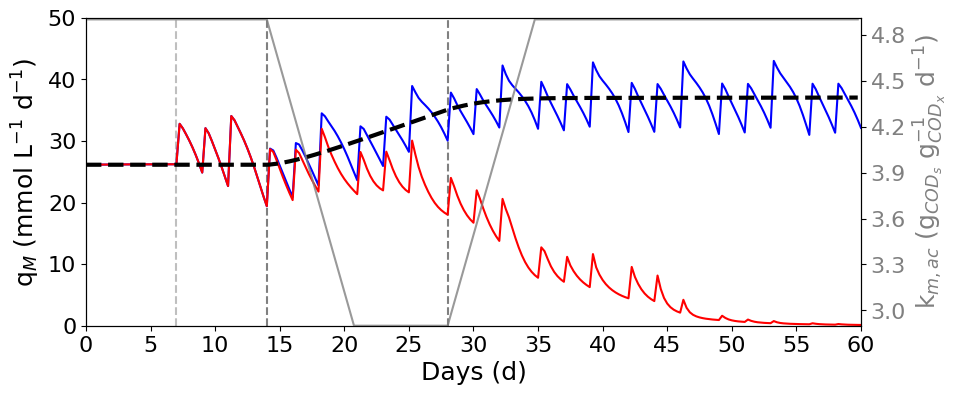}
 \end{center}
 \vspace{-0.3cm}
 \caption{Open-loop simulations with (red) and without (blue) the trapezoidal $k_{m,ac}$ reduction (secondary y-axis) over the diet \emph{transient}. The black thick dashed line is $q_{M,ref}$. Note: the $k_{m,ac}$ values are reported \emph{before} temperature correction as in the original ADM1 manuscript: \emph{after} correction, the initial value of 7.9 is reduced down to 4.7 g$_{COD_s}$ g$_{COD_x}^{-1}$ d$^{-1}$ (-40\%).}
 \label{fig:openloop_adm1}
\end{figure}

% Realistic tests
The more realistic tests (TEST4-6) were conducted to verify the considerations above. In such tests, the controller was strongly challenged not only by the higher structural complexity of the agri-AcoDM and the manual impulse feeding of $\mathbf{d_{ref}}$, but especially by the unknown temporary reduction over the \emph{transient} period of the methanogens' growth rate: indeed, this latter was artificially introduced to lead the open-loop feeding of the \emph{offline}-optimized diet to process failure, as shown in Figure \ref{fig:openloop_adm1}, as a result of a VFA accumulation.

\begin{figure}[!htb]
 \begin{center}
  \includegraphics[width=1\columnwidth]{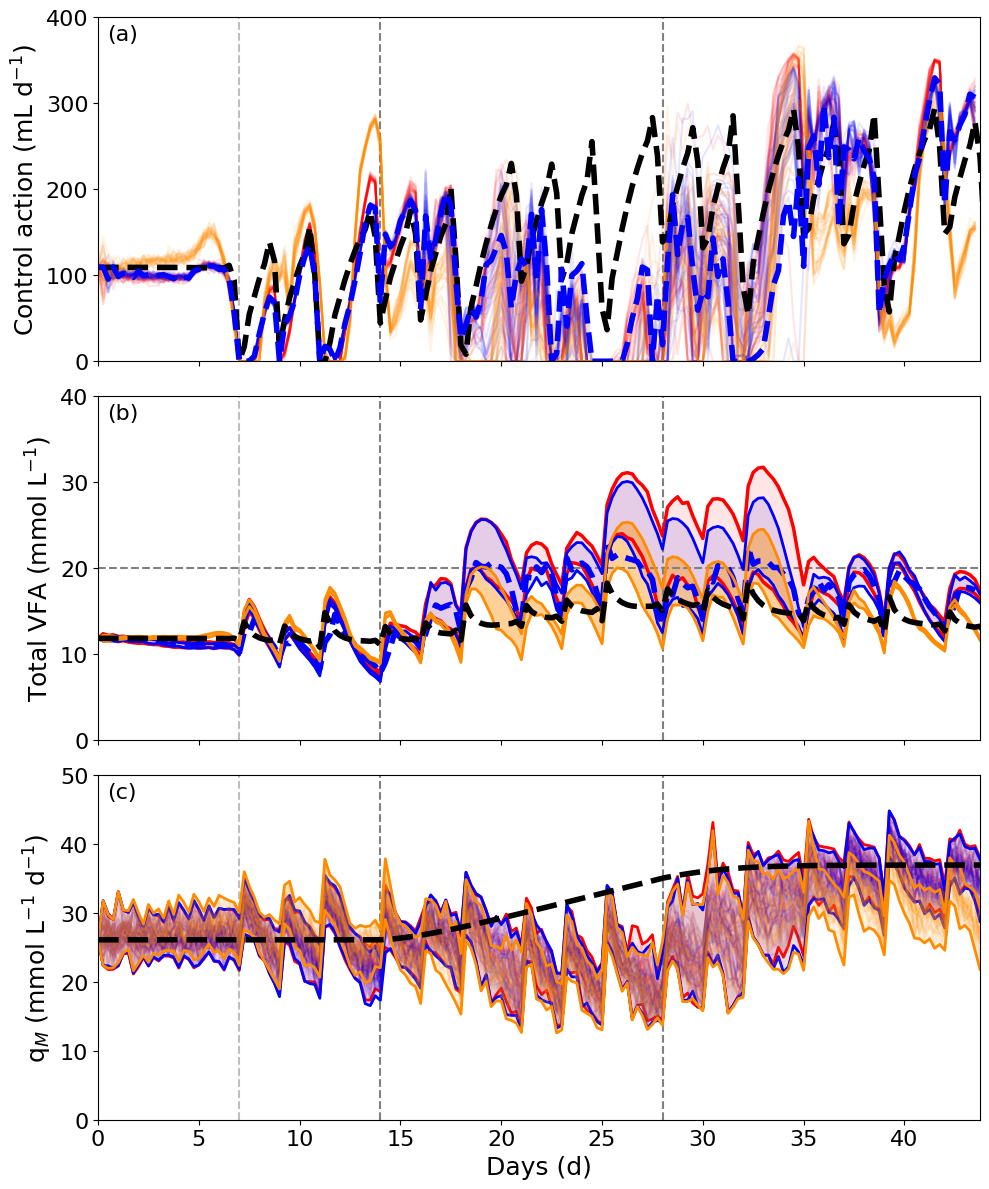}
 \end{center}
 \vspace{-0.3cm}
 \caption{Results from the Monte-Carlo analyses of TEST4 (red), TEST5 (orange) and TEST6 (blue). The black thick dashed line in (c) is $q_{M,ref}$. The thick dashed lines are: \emph{offline} (black) and \emph{online} (blue) $v^*$ (in (a)) and $\mathbf{z^*}$ (in (b)). The horizontal gray dashed line highlights the (soft) constraint on $S_2$.}
 \label{fig:closedloop_realistictests}
\end{figure}

\begin{figure}[!htb]
 \begin{center}
  \includegraphics[width=1\columnwidth]{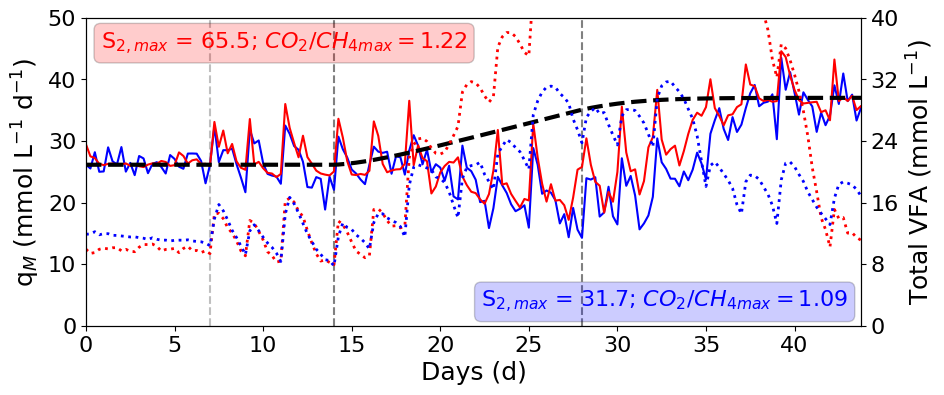}
 \end{center}
 \vspace{-0.3cm}
 \caption{Closed-loop simulations of \emph{classical} NMPC (i.e. one realization between the ones of TEST4, red) and 'override-PIs' (blue) with the trapezoidal $k_{m,ac}$ reduction over the diet \emph{transient}. The black thick dashed line is $q_{M,ref}$. The $q_M$ (primary y-axis) and $S_2$ (secondary y-axis) were reported with continuous and dotted lines respectively. Text boxes are colored as the tests they refer to.}
 \label{fig:closedloop_comparison}
\end{figure}

The Monte-Carlo results for these tests are depicted in Figure \ref{fig:closedloop_realistictests} to compare the \emph{classical} (TEST4, red color), the \emph{offline}- and \emph{online}-tube NMPC formulations (TEST5 and TEST6, orange and blue colors). Note that the three vertical dashed gray lines mark the start time of the impulsive $\mathbf{d_{ref}}$ and the start-end times of the $\mathbf{d_{ref}}$ ramp ($\sim$ \emph{transient} period). The weights $w_{y_{H_p}}, w_x$ and $w_u$ of the \emph{ancillary} cost function were set to 1, 1 and 0.1 respectively. Interestingly, the NMPC controller was able to deal with the disturbances and to avoid process failure in all its formulations. First of all, to reject the impulsive nature of the known $\mathbf{d_{ref}}$, the feeding of tomato sauce is delayed with respect to the instant of manual feeding of the other co-feedstocks: this is likely due to a prediction horizon that spans two consecutive impulse feeding instants. Furthermore, to reject the unknown reduction in the growth rate of the methanogens in the true plant, the control action $u$ over the \emph{transient} is much lower compared to the open-loop one and to the one suggested by the \emph{nominal} problem in the \emph{offline}-tube formulation ($\bm{\nu}^*$), with a consequent lag in the tracking of $q_{M,ref}$ (Figure \ref{fig:closedloop_realistictests}, days 22-28).
The results confirmed the considerations stated above for the preliminary tests: apparently, the tube-based formulation has a hard time to out-compete the \emph{classical} one when in the latter state constraints are tightened and slack variables are present, and tighter state tubes can be obtained only with a worsening of the $q_{M,ref}$ tracking performance. In addition, due to its higher flexibility in shaping $\mathbf{z}^*$ and beyond the values of the \emph{ancillary} cost function's weights, the \emph{online}-tube NMPC appears to be itself a good trade-off between the \emph{classical} formulation and the more 'conservative' \emph{offline}-tube one, balancing the better tracking performance of the first with the higher robustness/safety (i.e. lower $S_{2,max}$ and $(CO_2/CH_4)_{max}$) of the second one.

% Comparison with selector PI
Figure \ref{fig:closedloop_comparison} shows the comparison between one \emph{classical} NMPC and one 'override-PIs' closed-loop simulation: both controllers were able to deal with the unknown disturbance on $k_{m,ac}$, with a similar lag in the $q_{M,ref}$ tracking and without any particular benefit of the NMPC in the tracking performances. However, the NMPC seems to out-compete the 'override-PIs' in robustness, since (i) its control action is much less prone to saturation and 'bang-bang' behavior, and (ii) the $S_{2,max}$ and $(CO_2/CH_4)_{max}$ are significantly lower, indicating 'safer' operations (see the text boxes in Figure \ref{fig:closedloop_comparison}).

\section{Conclusions and future work}
\label{sec:Conclusions}
% Summary
An industrial-oriented application of a robust NMPC to the anaerobic co-digestion of agro-zootechnical feedstocks was presented. Different simulation tests were conducted both in 'simplified' and more realistic conditions, to allow for a comprehensive and fair comparison between \emph{classical} and tube-based NMPC formulations.

% Conclusion
Despite the higher ability of the tube-based NMPC to reject the presence of measurement noise and uncertainty in the model kinetic parameters, a trade-off between two competing control objectives was highlighted: thus, the tuning of the controller's hyper-parameters (i.e. the weights in the \emph{ancillary} cost function) is up to decision makers, which is whether to obtain either a better bio-methane flow rate tracking (performance) or lower total VFA concentrations i.e. 'safer' operations (robustness). The presence of slack variables in the \emph{classical} formulation and \emph{nominal} problems appeared to be beneficial to balance performance and robustness. The additional degree of freedom of the \emph{online}-tube formulation seems to have a similar effect, whereas the \emph{offline}-tube may be overly conservative. 

Compared to the 'override-PIs' scheme of \cite{carecci2024selector}, the NMPC was not able to improve the bio-methane flow rate tracking during the \emph{transient} period of diet change, but it resulted in more stable operations, enlightening, preliminarily, the benefits of a model-based over a conventional approach.

% Futher works
Overall, considering the realistic and strong sources of unknown disturbances, and the tough operative testing conditions (i.e. load and diet composition change in time), the results look promising: the main perspective of this study is thus the experimental validation of the NMPC on a small pilot-scale plant under similar operative conditions, including a monitoring configuration setup with hard and soft sensors. To achieve such goal, as previously stated, it is important to stress that $\mathbf{x}$ is not measurable \emph{online} in the industrial practice, so that the design of a state observer that estimates $\mathbf{x}$ from $\mathbf{y}$ \cite{Dewasme2019} is required for the actual implementation of the closed-loop and it is the first next step to be undertaken. 

Further works also include: (i) the comparison of the tube-based with other efficient robust MPC formulations (e.g. multi-stage scenario-based), and with a linear approach similar to the one proposed in \cite{GHANAVATI202187}; (ii) the extension of the current NMPC optimization problem to an economic framework, to penalize the use of expensive input feedstocks. Eventually, the ultimate goal is the extension of the present controller to a multiple-input multiple-output (MIMO) case, for its consequent validation in a full-scale plant equipped with screw-presses to automate the feeding of all the co-feedstocks (TS-rich ones too). 

With this prospects, the development of such a \emph{digital twin} could become a powerful tool to improve the techno-economic performance of bio-methane production from waste mixtures.

\section*{Acknowledgements}
This work was supported by A2A S.p.A. (Agripower Group), the European Union – Next-Generation EU (National Agritech Center) and the Italian Ministry of University and Research (National Recovery and Resilience Plan).

\bibliographystyle{abbrv}
\bibliography{NMPC}
\vspace{-20pt}\textcolor{white}{ } % Evita che il template sminki il margine sx dell'ultima riga

\end{document}